\documentclass[11pt]{article}
\usepackage[top=2.0cm, bottom=3.0cm, left=2.2cm, right=2.2cm]{geometry}
\usepackage{setspace} 
\usepackage[super,comma,sort&compress]{natbib}       
\usepackage{authblk} 
\usepackage{orcidlink}
\usepackage{float}
\usepackage{adjustbox}

\graphicspath{{./figures/}}

\usepackage{enumitem}
\usepackage{xspace}
\usepackage{fancyhdr}
\usepackage{afterpage}
\usepackage{comment}
\usepackage{listings}
\usepackage{makeidx}
\usepackage{longtable}
\usepackage{chemformula}
\usepackage[utf8]{inputenc}
\usepackage{comment}

\usepackage[dvips]{epsfig}
\usepackage{epstopdf}
\usepackage{graphicx}

\usepackage{booktabs}

\usepackage{mathtools}
\usepackage{amsmath}
\usepackage{amssymb}
\usepackage{bm}
\usepackage{amsfonts}
\usepackage{commath} 

\usepackage{url}

\newcommand{\kalpha}{\ensuremath{K_{\alpha12}}\xspace}

\newcommand{\mud}{\ensuremath{\mu D}\xspace}
\newcommand{\qmax}{\ensuremath{Q_{\mathrm{max}}}\xspace}

\newcommand{\qmin}{\ensuremath{Q_{\mathrm{min}}}\xspace}
\newcommand{\qbroad}{\ensuremath{Q_{\mathrm{broad}}}\xspace}
\newcommand{\qdamp}{\ensuremath{Q_{\mathrm{damp}}}\xspace}
\newcommand{\deltatwo}{\ensuremath{\delta_2}\xspace}

\newcommand{\rpoly}{\ensuremath{r_{poly}}\xspace}
\newcommand{\rw}{\ensuremath{R_w}\xspace}
\newcommand{\rapdf}{\textit{RA-PDF}\space}

\newcommand{\rmin}{\ensuremath{r_{\mathrm{min}}}\xspace}
\newcommand{\rmax}{\ensuremath{r_{\mathrm{max}}}\xspace}

\newcommand{\iq}{\ensuremath{I(Q)}\xspace}
\newcommand{\sq}{\ensuremath{S(Q)}\xspace}
\newcommand{\fq}{\ensuremath{F(Q)}\xspace}
\newcommand{\gr}{\ensuremath{G(r)}\xspace}

\newcommand{\q}{\ensuremath{Q}\xspace}
\newcommand{\ir}{\ensuremath{r}\xspace}

\newcommand{\tth}{\ensuremath{2\theta}\xspace}

\newcommand{\iaa}{\AA\ensuremath{^{-1}}\xspace}

\newcommand{\fig}[1]{Fig.~\ref{fig:#1}}

\newcommand{\sect}[1]{Section~\ref{sec:#1}}

\newcommand{\tabl}[1]{Table~\ref{table:#1}}

\newcommand{\adhoc}{\textit{ad hoc}\xspace}

\graphicspath{ {./figures/} }

\newcounter{saveenumi}

\newcommand{\be}{\begin{enumerate}}
\newcommand{\ee}{\end{enumerate}}
\newcommand{\bes}{\begin{enumerate}[wide, labelwidth=!, labelindent=0pt, label=\textbf{\textcolor{blue}{\arabic*}.}]}
\newcommand{\ees}{\end{enumerate}}

\newcommand{\getx}{{\sc PDFgetX3}\xspace}

\newcommand{\cmi}{{\sc Diffpy-CMI}\xspace}

\newcommand{\degree}{\ensuremath{^\circ}\xspace}

\newcommand{\tstep}{\ensuremath{t_{\mathrm{step}}}\xspace}
\newcommand{\deltarms}{\ensuremath{\Delta_{\mathrm{rms}}}\xspace}
\newcommand{\deltat}{\ensuremath{\Delta t}\xspace}
\newcommand{\mka}{Mo~\kalpha}   
\newcommand{\esv}{\ensuremath{V_e}\xspace}
\newcommand{\ttmin}{\ensuremath{2\theta_{\mathrm{min}}}\xspace}
\newcommand{\ttmax}{\ensuremath{2\theta_{\mathrm{max}}}\xspace}

\begin{document}  

\author[*,1,2\orcidlink{0000-0003-4989-5821}]{Till Schertenleib}
\author[2]{Daniel Schmuckler}
\author[2]{Yucong Chen}
\author[*,3]{Geng Bang Jin}
\author[1\orcidlink{0000-0002-8375-2341}]{Wendy~L. Queen}
\author[*,2\orcidlink{0000-0002-9734-4998}]{Simon~J.~L. Billinge}

\affil[1]{Laboratory for functional inorganic materials (LFIM),
Institut des Sciences et Ingénierie Chimiques,
École Polytechnique Fédérale de Lausanne (EPFL), Rue de l’Industrie 17,
CH-1950 Sion, Switzerland}
\affil[2]{Department of Applied Physics and Applied Mathematics, Columbia University,  New York, NY 10027, United States}
\affil[3]{X-ray Laboratory, 3M Corporate Research Analytical Laboratory, 3M Center, St. Paul, MN 55144, United States}

\title{Testing Protocols for Obtaining Reliable PDFs from Laboratory x-ray Sources Using \getx}

\maketitle
\doublespacing
\setlength{\parindent}{0pt}

\begin{abstract}
In this work, we explored data acquisition protocols and improved data reduction protocols using \getx to obtain reliable data for atomic pair distribution function (PDF) analysis from a laboratory-based Mo $K_\alpha$ x-ray source.  A variable counting scheme is described that preferentially counts in the high-angle region of the diffraction pattern. The effects on the resulting PDF are studied by varying the overall count time, the use of Soller slits, and limiting the out-of-plane divergence of the incident beam. The protocols are tested using an amorphous silica and a quartz sample. We present a Python script that allows the comparison of PDFs obtained from different instruments and different acquisition protocols and does automatic uncertainty estimation and error propagation. We also present a modification to the current \getx data corrections to take care of sample absorption, which was previously neglected in the use of that program for high-energy synchrotron x-ray data.  
\end{abstract}

\newpage
\section{Introduction}
  
The atomic pair distribution function (PDF) analysis of powder diffraction data is a widely used technique for studying the structure of nanomaterials \cite{keenCrystallographyCorrelatedDisorder2015, a.youngApplicationsPairDistribution2011, billingeNanoscaleStructuralOrder2008d, lindahlchristiansenThereNoPlace2020a}, amorphous materials \cite{wagnerDirectMethodsDetermination1978, benmoreReviewHighEnergyXRay2012} and liquids \cite{furukawaRadialDistributionCurves1962, fischerNeutronXrayDiffraction2005}, and increasingly to study nanostructures in crystals \cite{keenTotalScatteringPair2020, lin;prb05, beecherDirectObservationDynamic2016b}. 
It is currently most widely applied to diffraction experiments carried out at x-ray synchrotron sources and neutron sources which are large-scale national and international shared facilities.  
However, data useful for PDF analysis may also be obtained from laboratory-based instruments that have relatively short wavelength sources such as Mo or Ag $K_{\alpha}$ x-rays \cite{egami;b;utbp12,dykhn;phmr11,thomaePushingDataQuality2019, tsymbarenkoQuickRobustPDF2022, prinzHardXraybasedTechniques2020, irvingAdvantagesCurvedImage2021, confalonieriComparisonTotalScattering2015}. 
These types of PDFs are set to grow in popularity because of access challenges at the national facilities, and the emergence of specialized instruments for PDF analysis on the laboratory x-ray diffractometer market \cite{D8ADVANCE, D8DISCOVER, EmpyreanXRDInstrument, Stadi, Rigaku, antonpaar}. 
Counting statistics in raw diffraction data are extremely important in order to obtain a reliable PDF. This poses a challenge for laboratory instruments, since their x-ray flux is of course much lower compared to synchrotron sources, and much longer acquisition time is required. Specialized instruments have been developed with multi-moduli detectors \cite{thomaePushingDataQuality2019} or curve image plate detectors \cite{irvingAdvantagesCurvedImage2021} that allow covering a larger angular range simultaneously, and satisfactory counting statistics up to 20~\iaa can be achieved in less than 6 hours using monochromatic Ag $K_{\alpha1}$ radiation.
Despite offering a lower accessible \q-range, Mo $K_{\alpha12}$ ($\alpha1,\alpha2$ mixture) sources have a number of advantages over Ag sources, such as higher flux and longer source life, and can also yield PDFs with sufficient real-space resolution for many scientific problems \cite{dykhn;phmr11,tsymbarenkoQuickRobustPDF2022}. 

Many research institutes are equipped with single crystal x-ray diffractometers with \mka sources and large area 2D~detectors. These instruments are perfectly suitable for obtaining reliable PDFs. For example, \cite{tsymbarenkoQuickRobustPDF2022} conducted a detailed study, comparing PDFs obtained from different x-ray sources (Mo, Ag) and different diffractometers (single crystal diffractometers Bruker D8 QUEST, Bruker D8 Advance, and powder diffractometer STOE STADI P). Taking advantage of the high flux of a Mo~$K_{\alpha12}$ and high signal-to-noise ratio achievable using a large 2D detector, they were able to obtain PDFs of polycrystalline powders comparable to those obtained from a synchrotron measurement. With just 90 minutes of acquisition time, they could obtain high-quality diffraction data with a \qmax=16~\iaa, yielding PDFs with accurate peak positions up to 40~\AA, and after $K_{\alpha2}$ stripping, even up to 80 \AA. \cite{tsymbarenkoQuickRobustPDF2022} also presented software that allows streamlining the integration of each 2D snapshot and merging into 1D diffraction patterns for subsequent data reduction to \gr.

Though a number of groups have explored the use of specialized and standard laboratory diffractometers for obtaining quantitative PDFs, less attention has been given to standard analysis protocols and benchmarking to synchrotron data. 
This paper deals with protocols for acquiring and analyzing data from such lab instruments to obtain accurate PDFs using the latest-generation data reduction software packages.

A challenge in PDF analysis is to obtain a high-quality PDF from raw diffraction data since the quality of the resulting PDF depends not only on the choices made during data acquisition but also on how the data reduction is carried out from the raw data to the PDF. 
Two contrasting approaches to data reduction are often followed.
One makes corrections for experimental effects by explicitly calculating them from the understood physics and geometry and making algebraic corrections to the data to account for those effects \cite{egami;b;utbp12}.  
The problem with this approach is that many of the experimental factors and their contributions to the scattering signal are only approximately known.
It has been shown that making certain \adhoc corrections that are justified by the physics of the situation and the fact that reasonable assumptions about the experiment hold can result in PDFs that are equivalent to those obtained using the physics-based data reduction approach \cite{juhas;jac13}.  
In fact, the \adhoc correction approach can be superior in some cases, for example, when subtracting large backgrounds to extract weak signals of interest, when it is not guaranteed that the foreground and background measurement geometries are exactly reproduced. 
This allowed signals to be extracted from nanosized organic drug particles in aqueous solution at concentrations of 0.2 wt\% \cite{terba;nan15} and even the observation of different hydrogen bonding networks in differently amorphized polymeric samples \cite{terbanLocalStructuralEffects2020}.

Here, we explore how to use the \adhoc PDF data analysis program, \getx, to obtain reliable PDFs from a Bruker D8 Discover laboratory instrument. Effects of beam divergence and counting time were tested by measuring a NIST quartz standard and an amorphous silica sample with different acquisition protocols that vary in the overall counting time, the use of axial Soller slits, and the use of masks to limit additional beam divergence. We provide a Python script for PDF analysis, uncertainty estimation and error propagation that allows a quantitative comparison of PDFs obtained from different protocols and allows bench-marking to synchrotron data.
We discuss challenges in data reduction when using low-energy x-ray sources in laboratory diffractometers.
Some of the approximations inherent in the use of \getx, which was developed for high-energy synchrotron experiments, are not necessarily valid for the lab data and additional corrections for these effects are discussed.

\section{\emph{Ad hoc} analysis procedure in \getx}

The \adhoc approach used in \getx for obtaining the PDF from powder diffraction data is described in detail in \cite{juhas;jac13}.
It was shown \cite{juhas;jac13}, in the case of high energy synchrotron data, that this approach resulted in PDFs that were identical to those obtained using a conventional, physics-based, analysis using the {\sc PDFgetX2} program \cite{qiu;jac04i}. 

We summarize the pertinent aspects of the \getx method here.
In general, multiplicative and additive/subtractive corrections need to be applied to a measured diffraction pattern to turn it into the relevant scattering function, \sq. 
Multiplicative corrections deal with things such as sample absorption and geometric effects that affect the illuminated sample volume \cite{egami;b;utbp12}. 
Additive corrections deal with effects of multiple scattering, fluorescence, Compton scattering, and so on \cite{egami;b;utbp12}.
In many cases we don't care about the absolute value of the overall scale factor since that parameter can be refined to fit structural models and doesn't affect the refined values of structural parameters \cite{juhas;jac13}.  
If we don't care about an overall scale-factor, it is only the \q-dependence of the multiplicative and additive aberrations that is important, \q being the magnitude of the scattering vector, $Q=\frac{4\pi \sin\theta}{\lambda}$, where $\theta$ is half the scattering angle and $\lambda$ is the x-ray wavelength. 
For the case of high-energy synchrotron radiation experiments, in most cases, the absorption and sample volume effects give very weak \q~dependencies and can be neglected.
On the other hand, especially for \rapdf (rapid acquisition PDF) experiments \cite{chupasRapidacquisitionPairDistribution2003f} carried out with large area 2D detectors, the additive effects are significant.
The \getx approach uses a polynomial function to fit the \q-dependence of these signals and removes it by subtraction.
The flexibility of the polynomial is very important.
It needs to be flexible enough to remove slowly varying incoherent backgrounds that do not contain structural information, but not to remove any meaningful structural signals.
In \getx the flexibility is determined by the order of the polynomial function that is fitted, where a higher order results in a more flexible fitting curve.
There is no structural signal in \sq that has a frequency lower than $2\pi/r_{nn}$, where $r_{nn}$ is the inter-atomic distance of the nearest neighbor peak in the PDF, and so limiting the order of the polynomial to some value such that no frequencies higher than this can be removed ensures no loss of structural information.
The polynomial order, and thus the fitting flexibility, is controlled by the \rpoly parameter in the program.
After removal of the additive aberrations, the signal is normalized to obtain \sq by satisfying its known asymptotic behavior, so that its low- and high-\q behavior is correct.

In principle, we could use the same approach for data from laboratory instruments, but the assumption of negligible x-ray absorption by the sample is no longer valid in general due to the longer wavelength x-rays. 
This both increases the importance of the absorption effects overall but additionally introduces significant \q~dependencies to them because data have to be measured over wide angular ranges to obtain reasonable values of \qmax, and the effective illuminated volume of the sample becomes significantly angle-dependent \cite{egami;b;utbp12}.
For quantitatively accurate PDFs, new steps need to be added to the \getx algorithm to account for this issue.

\section{Methods}

We carried out a set of \mka experiments with quartz and silica glass samples.  
These were chosen because of their relatively moderate absorbing power for \mka radiation. For a path length $l=1.5$~mm, $\mu l = 1.11$ is estimated where $\mu$ is the linear x-ray attenuation coefficient. Typically, values below 1 are unproblematic, whereas larger values might become an issue. We also point out, as we will discuss later, that there is a \tth\ dependence to the x-ray attenuation in a typical capillary geometry.
The results can be extended to more absorbing samples later.
Quartz was chosen because it is crystalline and it is therefore possible to model it quantitatively.
Silica glass was chosen because, on the other hand, it is amorphous and contains only diffuse scattering.
Such data with weak signals can be more challenging for acquiring data suitable for PDF analysis and so the combination of a crystalline and an amorphous test sample covers both cases.
To benchmark the quality of the results, we compare the resulting PDFs to expected results from quartz structural models and against PDFs obtained from the same samples measured using high-energy synchrotron x-rays using a standard \rapdf approach.

\subsection{Laboratory Experiments}\label{sec:lab_experiment}

\mka data was obtained with a Bruker~D8~Discover diffractometer equipped with a line focus \mka radiation source (average $\lambda=0.7107$~\AA), and an EIGER2~R~500K detector operated in 1D mode. The transmission PDF configuration includes a focusing Goebel mirror, a 1.0~mm divergence slit, a 2.5\degree axial Soller slit, a scattering guard after the source, a capillary spinner on a centric Eulerian cradle, and an optional 2.5\degree axial Soller slit before the detector.   The source-to-sample distance is 200~mm, and the detector-to-sample distance is 198.5~mm. X-ray generator settings of 40~kV and 40~mA were employed.
Whilst flat plate geometries such as Bragg-Brentano can give increases in detected scattered flux, they can introduce data correction challenges in PDF because of the difficulty in estimating the illuminated sample volume \cite{egami;b;utbp12}.
To avoid this, and at the expense of a loss in measured intensity, we chose to use the capillary geometry, which is the least susceptible to these uncertainties, with the powdered samples contained in polyimide capillaries with an internal diameter (ID), $d_c = 1.5$~mm, and of approximate length~50~mm. The choice of $d_c$ is important and depends on the absorbing properties of the samples.  
Guidelines for choosing $d_c$ are described below in \sect{abs} and \sect{recommendations}.
\begin{figure}
    \centering
    \includegraphics[width=0.5 \textwidth]{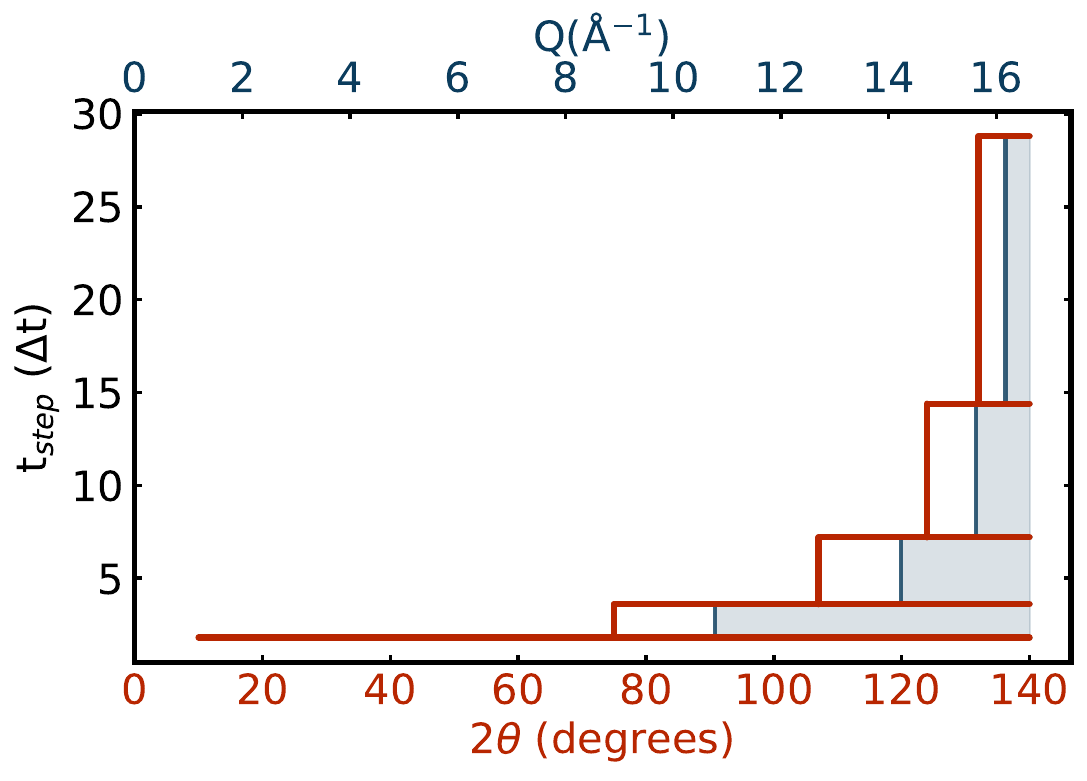}
    \caption{The staircase-count-time, SCT, protocol used for our experiments. The red lines indicate the range of \tth\ for each step, and the blue lines indicate the range of \q\ for each step, with total time per point \tstep\ which are plotted in units of the fixed count-time interval, \deltat = 1.8~seconds in this case. \deltat\ may be varied from experiment to experiment to change the overall measurement time but this is done without changing other aspects of the protocol.}
    \label{fig:tstep}
\end{figure}
Scans were conducted with an angular step size of 0.025\degree~using an EIGER2 detector operated in Continuous 1D (Line beam geometry) mode with 69~mm (equatorial) $\times \, 24$~mm (axial) opening, which is a fast continuous scan mode with a standard over-travel by half of the detector opening (BRUKER lab Report XRD 90 and Application Note 618) \cite{eiger2}. The step size was chosen based on our estimate that the beam height on the detector was around 0.5~mm, which corresponds to an angular height of 0.15\degree; we chose an angular step size of 0.025\degree as a convenient round number less than 0.15/5 for the step size to ensure sufficient sampling of each peak.
During a continuous 1D scan, pixels are binned into horizontal strips, each parallel to the axial direction at a different \tth angle. All the strips sweep through the entire \tth range, passing through each point in \tth one strip at a time.
A special low-angle scan was conducted using the same detector but operated in no-over-or-under-travel standard continuous 1D mode.  
This allowed us to collect the lowest angle data while avoiding a collision between the detector arm and the goniometer floor/platform which would have happened if we had extended the standard continuous scans all the way to the lowest angles.  
This was not required for the optimal protocol but was a detail of specifics of the diffractometer platform itself.\\
\\
\begin{table}
\caption{The SCT protocol used for our experiments. $\tstep^i = c_i \deltat$ and $\ttmin^i$ are the count time per point and the minimum \tth angle for the scan of the $i^{th}$ step.  In each case, the maximum \tth value was 140~\degree, and the step size was $\Delta 2\theta = 0.025\degree$.
In addition to the staircase indicated in the \fig{tstep} a low-angle scan in a different mode was collected from -4\degree to 15\degree to get all the low-\q region to the edge of the beamstop whilst avoiding a diffractometer collision. $N_p$ is the number of points in the step.
}
\label{table:staircase}
\begin{center}
\begin{tabular}{lrrrrr}
  \toprule
  Scan & $c^i$ & $\tstep^i$ & \ttmin (\degree) &\ttmax (\degree)& $N_p$\\
  \midrule
  0  & ~ & 2 & -4&  15\footnote{Special low-angle scan to avoid a collision of the diffractomter, please see text for details.}& 761\\
  1  & 1 & $1*\deltat$ & 10&  140& 5201\\
  2  & 2 & $2*\deltat$ & 75&  140& 2601\\
  3  & 4 & $4*\deltat$ & 107& 140& 1321\\
  4  & 8 & $8*\deltat$ & 124& 140& 641\\
  5  & 16& $16*\deltat$ & 132& 140& 321\\
  \bottomrule
\end{tabular}
\end{center}
\end{table}

In order to increase the counting statistics in the high-Q region, we employed a staircase-counting-time (SCT) protocol. Our SCT protocol is a variation of widely used variable counting scheme (VCS) protocols as, for example, outlined by Bruker and Panalytical (Panlytical application note 20130107). In our case, we conducted five scans ranging from \ttmin to \ttmax=140\degree, where in each scan, the counting time per step is doubled compared to the previous scan (see Figure \ref{fig:tstep} and Table \ref{table:staircase}). Raw XRD data from each SCT scan were integrated from the 2D detector, normalized in counts/second (cps), accumulated into one data set, and then exported as a .xy file by use of the Bruker DIFFRAC.EVA software.


To assess different collection strategies and instrument configurations we carried out an extensive set of different measurement protocols.
They all used the SCT protocol with a value of \deltat shown in \tabl{acquisition_protocols}, which is scaled in each STC scan at higher Q as shown in Table \ref{table:staircase}.\\

\begin{table}
\begin{center}
    \caption{Data collection protocols for the \mka experiments. 
    A staircase-counting-time (SCT) protocol was used, described in \sect{lab_experiment}. $\tth_{max}$ is always set to 140\degree and five steps in the stair were measured following the SCT protocol described in the text. \deltat is the count-time interval described above. The beam-height on the detector was 0.5~mm corresponding to a solid angle of~0.15\degree, and we used a step size of~0.025\degree  resultin in a total elapsed time for each experiment reproduced here.  The secondary beam Soller slit was a 2.5\degree slit where Y and N indicate whether the slit was installed or not, respectively. The illumination length refers to the length of the capillary sample that was exposed to the beam.}
    \label{table:acquisition_protocols}
    \scriptsize
    \begin{tabular}{lccccc}
    \toprule
       Protocol  & \deltat (s) & Elapsed time (hours) & Soller & Illumination length (mm) \\
    \midrule
        P1 & 0.1 & 1.5 & N & 25 \\
        & 0.3 & 4.5 & N & 25 \\
        & 0.9 & 12.75 & N & 25 \\
        P2  & 0.3 & 4.5 & Y & 25 \\
          & 0.9 & 12.75 & Y & 25 \\
          & 1.8 & 25.5 & Y & 25 \\
        P3 & 0.1 & 1.5 & N & 15 \\
           & 0.3 & 4.5 & N & 15 \\
    \bottomrule
    \end{tabular}
\end{center}
\end{table}

We carried out data collection without a Soller slit in the diffraction path (P1), with a 2.5\degree~Soller slit (P2), and reduced length of the sample capillary that was illuminated by the beam (P3). 
The latter was accomplished by cutting the edges of the incident beam using lead tape upstream of the sample to limit the length of the incident beam on the sample.
This led to a set of three acquisition protocols, each with various \deltat times, that are summarized in detail in \tabl{acquisition_protocols}.

A background measurement was performed with an empty polyimide capillary, using the same experimental configuration and measuring conditions, to account for the signal produced by the capillary material, air-scattering, or by any other source different from the sample itself.   

The merged XRD data were converted to \gr using \getx \cite{juhas;jac13} within xPDFsuite \cite{yang;arxiv14}. 
A $\qmax = 16.6$~\iaa was used in the Fourier transform from \fq to \gr. 
The Fourier transform was carried out on a grid of $\Delta r = 0.01$~\AA\ and also on the Nyquist-Shannon (NS) grid ($\Delta r = \frac{\pi}{\qmax} = 0.189$~\AA) \cite{farro;prb11}.
The former grid gives a smoother PDF which is easier to interpret visually whilst the latter results in data points in the PDF that are as statistically uncorrelated as possible without loss of information, and therefore results in better estimates of uncertainties during model fitting.

\subsection{Synchrotron Experiments}

Total scattering experiments were carried out on at the 28-ID-1 beamline at the National Synchrotron Light Source II (NSLS-II) at Brookhaven National Laboratory. A 2D Perkin-Elmer amorphous silicon detector ($2048\times2048$~pixels, $200\times 200$~micron pixel size) was placed perpendicular to the incident beam path 205~mm downstream of the sample. 
Samples were fine powders loaded in Kapton tubes with an inner diameter of 1~mm. 
A bent Laue monochromator was used to select the photon energy. The incident wavelength was $\lambda=0.1665$~\AA.
The instrument geometry was calibrated using a polycrystalline nickel standard using the pyFAI program \cite{ashio;jac15}. 
Data were collected for 15~minutes by summing 4,500 0.2~second frame exposures. 
Dark frames were collected before the actual exposures by exposing the detector in the same way but with the shutter closed.
These were subtracted from the bright images to remove detector dark current effects.
2D diffraction images were masked, and the raw data were integrated along arcs of constant diffraction angle using pyFAI with full pixel splitting, solid angle correction off, and 15000 radial bins to produce 1D~powder diffractograms.
Total scattering data from an empty 1~mm Kapton (polyimide) tube was collected to determine the background intensity due to the sample container and any air scattering.
The PDFs, \gr, were obtained following standard data reduction protocols described in detail elsewhere \cite{juhas;jac13} using~\getx within xPDFsuite \cite{yang;arxiv14}. The background signal was scaled and subtracted as part of this process in \getx. 

\qmin = 0.8 and \qmax = 26.0~\iaa were used in the Fourier transform that takes \fq to \gr. The Fourier transform was done onto a grid of $\Delta r = 0.01$~\AA~and also on the NS grid ($\Delta r = \frac{\pi}{\qmax} = 0.12$~\AA)\cite{farro;prb11}. 





\subsection{PDF Modelling}

Structure modeling was done using \cmi\cite{juhas;aca15}. The python script for carrying out fitting is available in the supplementary information. The structure model of $\alpha$-quartz was taken from the crystallographic open database (ID 1011097). The parameters \qdamp, \qbroad, the correlated motion parameter \deltatwo, the lattice parameters $a$ and $b$, the atomic displacement parameters (ADPs), the atomic fractional coordinates, $X$, $Y$ and $Z$, and a scaling variable $s_1$, were refined. The lattice parameters, ADPs, and atomic positions were constrained to satisfy the symmetries of the space group (P3\textsubscript{1}21). All the ADPs were initialized with a value of 0.01~\AA$^2$. \deltatwo was initialized with a value of 2.333, \qbroad and \qdamp were initialized with 0.0. Other parameters were initialized with their values in the CIF file for the first refinement, but to increase computational efficiency, subsequent refinements used the outputs of previous refinements to initialize the variables. Refinements were carried out on the Nyquist-Shannon (NS) grid, where the data were sampled onto the NS grid using the Whittaker-Shannon interpolation \cite{whittakerXVIIIOnFunctionsWhich1915, shannonMathematicalTheoryCommunication1948}. 
The PDFs were fitted between \rmin= 1.0~\AA~and \rmax= 40.0~\AA. 
It was previously shown that in that region, the use of a mixed \mka does not lead to significant loss of resolution and erroneous peak positions \cite{tsymbarenkoQuickRobustPDF2022}.

The refinements were done with an error propagation, with initial standard errors assigned to the points in the PDF data by calculating the root-mean-square ($rms$) of the fluctuations in the measured PDFs between $300<r<400$~\AA~(see section \ref{sec:statistics} for justification for this).
This results in uncertainties on the refined parameters that are shown in parentheses in~\tabl{refinement_results}.  This process is expected to result in statistically reasonable uncertainty estimates when the data is in the count-rate limited regime that applies to the laboratory data.  This does not apply to the synchrotron data where the high counting statistics mean that the fluctuations in the PDF at high-\ir are dominated by systematic errors. In this case, the estimated uncertainties shown in the parentheses result in a significant underestimation of actual uncertainties, though they are reproduced in the table for completeness.

\section{Results}
\subsection{Comparison of Lab and synchrotron data}

\begin{figure}
    \centering
    \includegraphics[width=0.6 \textwidth]{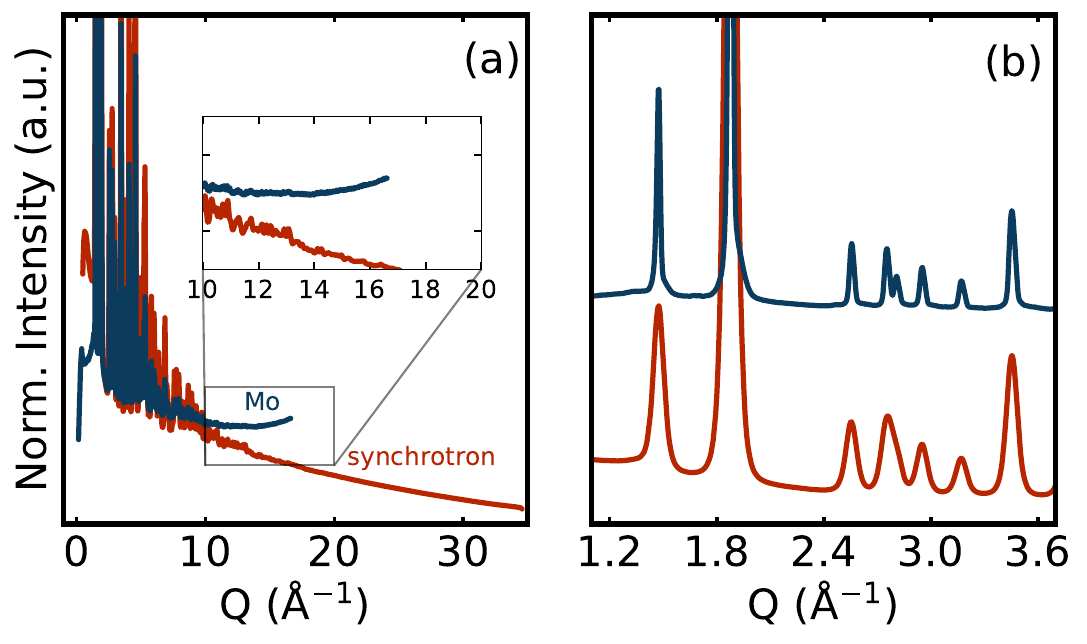}
    \caption{Comparison of the measured intensities from a quartz measurement using \rapdf synchrotron (red) and a \mka lab (blue) data.  The lab data were collected with protocol P2 and $\deltat = 1.8$~s
    (a) The data are plotted over the entire \q-range measured in each case.  The inset shows the intermediate range of~\q on an expanded scale. 
    (b) The low-\q range for the same data is shown on an expanded scale.
    In both plots the synchrotron and Mo data have been scaled to 1 by dividing by the maximum peak such that the amplitude of the low-\ir peaks in \gr are approximately equal in the two curves.}
    \label{fig:demo_abs_corr}
\end{figure}

We start by making a direct comparison of the 1D diffraction patterns from quartz obtained from the lab instrument and the synchrotron (\fig{demo_abs_corr}).
For the lab data, we consider the data from the measurement protocol P2 with \deltat = 1.8~s, the 25.5~h measurement with Soller slits in place.
 
The Mo x-rays ($\lambda=0.7107$~\AA) give us an achievable $\qmax = 16.6$~\iaa after measuring up to the maximum angle of $2\theta = 140\degree$, whereas the synchrotron data give us a much larger accessible Q range ($\lambda=0.1665$~\AA). 
This will limit the real-space resolution of the PDF from the \mka measurement.
We observe a stronger upturn in the \mka measurement with increasing-\q that is not present in the synchrotron data. This is shown on an expanded scale in the inset (\fig{demo_abs_corr}a). We hypothesize that this upturn is, in part, a result of sample absorption which will be weaker in the backscattering region  ($2\theta > 90 \degree$) than the forward scattering and therefore lead to a relatively stronger signal with increasing-\q. 
We explore this effect more quantitatively in \sect{abs}.
The low-\q region of the diffraction pattern is shown on an expanded scale in \fig{demo_abs_corr}(b).  
Since the \q-space resolution of the \mka data is higher than that of the \rapdf synchrotron data, this results in a structural signal in \gr that extends to higher-\ir \cite{egami;b;utbp12, proffenStructuralAnalysisComplex2009, tsymbarenkoQuickRobustPDF2022}.
\begin{figure}
    \centering
    \includegraphics[width=0.5\textwidth]{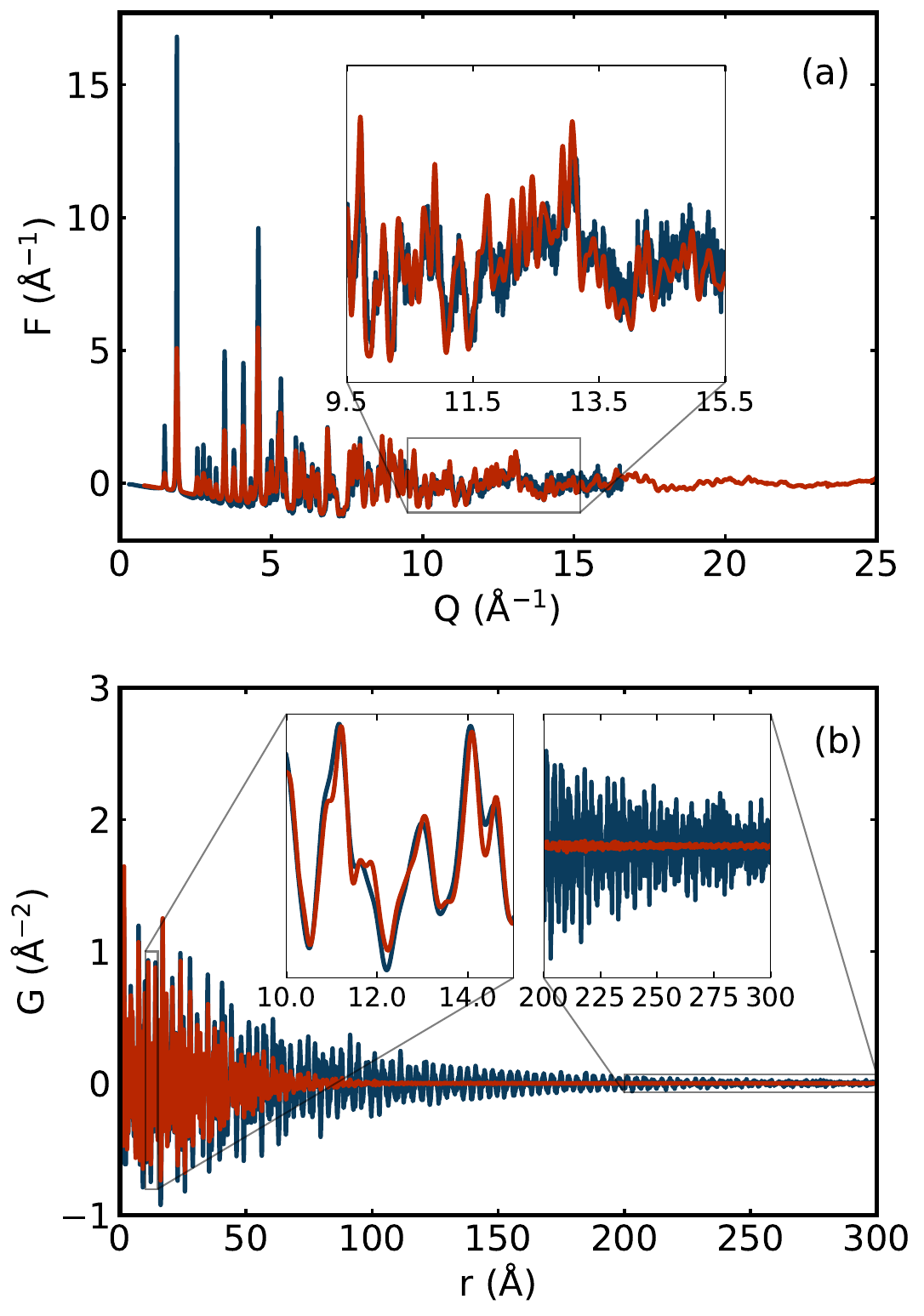}
     \caption{ Comparison of (a) \fq and (b) \gr functions of crystalline quartz collected via a synchrotron measurement (red) using the \rapdf method (total measurement time 15~minutes) and the \mka lab measurement (blue), collected with protocol P2 and $\tstep=1.8$~s (total time 25.5 hours). The insets show regions of the curves on an expanded scale. The synchrotron \fq and \gr were re-scaled by a factor of 0.85 to compare the two curves. This factor was chosen by making the \gr functions scale onto each other in the region $0<r<15$~\AA~(see insets in a and b)} 
     \label{fig:syncvsmo}
\end{figure}
\fig{syncvsmo} shows the corresponding \fq and \gr where, as before, the synchrotron data are shown in red and the \mka in blue. We see that the shape of the baseline of the synchrotron and \mka \fq align quite well, as \getx removes slowly varying changes in Q. As a result, we no longer see the upturn at high-\q in the \mka data.
However, as we argue in \sect{abs}, the \getx algorithm is leaving an uncorrected multiplicative correction due to absorption. For high incident x-ray energy synchrotron measurements (typically $0.15 < \lambda < 0.25$~\AA), neglecting this is a reasonable approximation. However, for \mka this is no longer true in general.
Such an uncorrected absorption effect would lead to diffraction intensities that are overestimated in the low-\q region and underestimated in the high-\q region of \fq in the \mka data compared to the \rapdf data.   

An incorrectly handled adsorption correction will produce changes to peak widths in \gr in the \mka data \cite{tobyAccuracyPairDistribution1992}.
In the left inset to \fig{syncvsmo}(b), we show a low-\ir region of the  PDFs on an expanded scale and, indeed, the PDF peaks from the \rapdf data are narrower. However, this is also because of the larger \qmax of the synchrotron data, and so the effect of the uncorrected absorption correction is not directly visible.
When structural models are fit to the data, it is expected to result in overestimated ADPs in resulting model fits. This is discussed in \sect{mult_corr}.

On the other hand, as expected due to the higher \q-space resolution, the structural signal in the \mka \gr extends further in~\ir than does the synchrotron PDF in red.
The signal in the synchrotron data dies out by about $r = 80$~\AA~and is replaced by noise, compared to the \mka \gr that clearly extends to above $r = 200$~\AA.  
High \q-resolution measurements may be made at synchrotrons using different setups \cite{fitchHighResolutionPowder2004}, but these experiments are slow, and fitting of models to PDFs are rarely done beyond an $\rmax = 40$ or 50~\AA~and the \rapdf setup \cite{chupa;iucrcpd03} is predominantly preferred.

The signal/noise level in the synchrotron data is much better than from the lab data despite the collection time being almost $50 \times$ less. This is clearly evident in the right-hand inset in \fig{syncvsmo}(b) that shows both PDF curves in the region from $250<r<300$~\AA. There is no structural signal left in this region in either the synchrotron or the \mka data and so the ripples in the PDF are completely coming from noise in the measurement.
The much smaller amplitude noise signal of the red curve clearly shows that in a 15~minute synchrotron scan, the noise levels are still much lower than in a 12.75~hour \mka scan. This can also be seen directly in \fq, where the diffraction signal is better resolved in the synchrotron data compared to the \mka data (see \fig{syncvsmo}(a), inset).
The high throughput of synchrotron measurements is a clear advantage for rapid experiments.

Despite this comparison, which appears to reflect negatively on the \mka data, it is possible to obtain reliable PDFs from laboratory sources with protocols that require much shorter acquisition protocols \cite{thomaePushingDataQuality2019, tsymbarenkoQuickRobustPDF2022, prinzHardXraybasedTechniques2020, irvingAdvantagesCurvedImage2021}. There are various instrumental settings that can ultimately determine the counting statistics. Below we explore optimizing counting statistics and systematically explore the effects of instrumental settings on the resulting PDFs.

\begin{figure}
    \centering
    \includegraphics[width=0.6\textwidth]{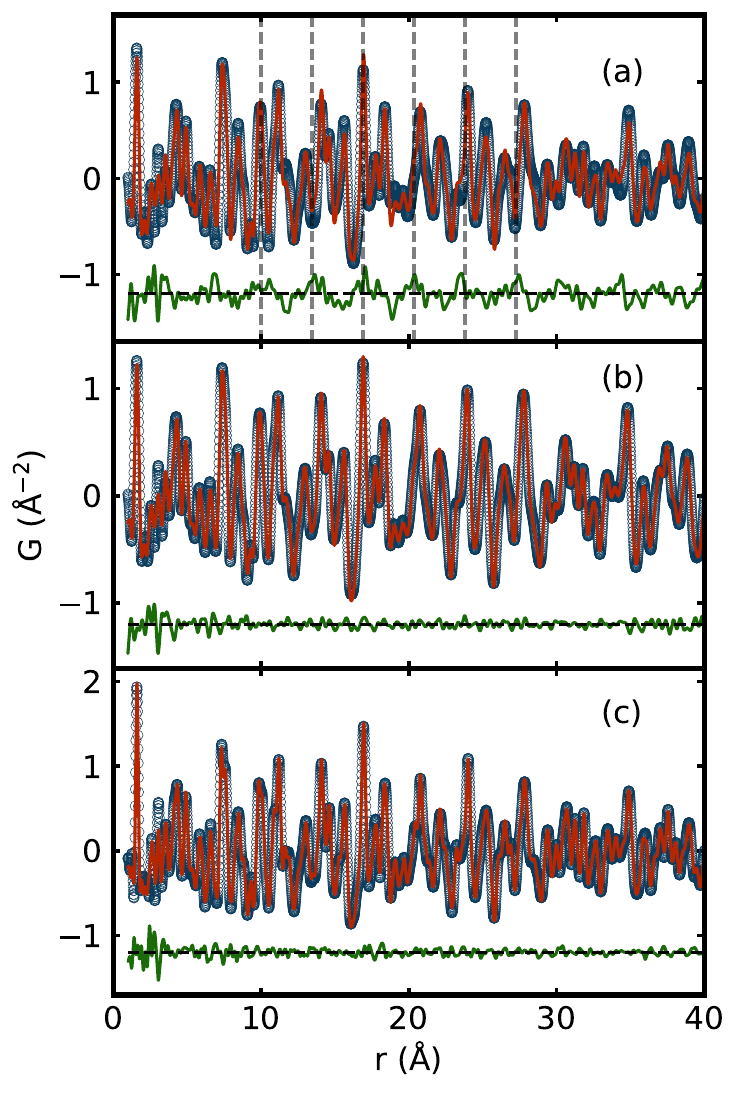}
    \caption{Comparison of best-fit calculated and measured PDFs for quartz. The measured PDF is shown as blue circles and the best-fit PDF as a solid red line. 
    The difference between the two is shown offset below. The data are (a) with no Soller slit (protocol P1) and 4.5~h measurement time, (b) with Soller slit (protocol P2) and 25.5~h measurement time, and (c) the synchrotron data. The vertical dotted lines in (a) highlight the oscillating singnal in the difference curve with an interval of 3.45~\AA. }\label{fig:pdf_fit_quartz}
\end{figure}

We now assess the quality of model fits to the synchrotron and \mka data.  
\fig{pdf_fit_quartz}(a) and (b) show fits to the \mka data without and with Soller slits, respectively, and the synchrotron data are shown in \fig{pdf_fit_quartz}(c).  
The best agreement between model and data is evidently the \mka data measured with the Soller slit in place (\fig{pdf_fit_quartz}(b)) and the synchrotron data \fig{pdf_fit_quartz}(c)).  

The refined parameters are reproduced in \tabl{refinement_results} where we compare the P2(1.8) and the synchrotron columns.
\begin{table}
\begin{center}
\caption{Results of the quartz model refinements to the data measured with selected protocols.  The protocols are written as P$N$(\deltat), where $N$ is the number of the protocol from \tabl{acquisition_protocols}.  Abs indicates the dataset that has had an absorption correction applied.
The uncertainties are one standard deviation. The estimated errors on the input data are computed by evaluation the root-mean-square fluctuations of the PDF in the region from $300 \le r \le 400$~\AA\ on the Nyquist-Shannon grid.  The justification for this is in the text.}
\label{table:refinement_results}
\begin{adjustbox}{angle=-90}
\begin{tabular}{llllllll}
\toprule
Parameter & P1(0.1) & P1(0.3) & P2(1.8) & P2(1.8) Abs & P2(0.3) & P3(0.3) & synchrotron \\
\midrule
       scale &                     0.2467(11) &                      0.2488(6) &                        0.2649(6) &                               0.2898(8) &                      0.2615(14) &                 0.2566(25) &                      0.246568(35) \\
    \qdamp &                    0.03359(26) &                    0.03405(15) &                      0.01880(20) &                             0.02093(22) &                       0.0193(5) &                  0.0317(6) &                       0.034291(8) \\
   \qbroad &                      0.0263(8) &                      0.0242(5) &                      0.02890(34) &                               0.0270(5) &                       0.0272(8) &                 0.0177(18) &                      0.013671(31) \\
   $\delta _2$ &                        2.24(6) &                        2.25(5) &                        2.224(11) &                                2.24(14) &                        2.25(11) &                   2.24(31) &                         2.0571(9) \\
        a &                    4.91384(14) &                   4.913910(35) &                      4.90482(10) &                             4.91055(13) &                     4.90561(24) &                  4.9063(4) &                       4.916540(5) \\
        c &                    5.41107(24) &                    5.41177(14) &                      5.39863(18) &                             5.40225(23) &                       5.3974(4) &                  5.4005(8) &                      5.409109(10) \\
    Si(U22) &                    0.00441(12) &                     0.00450(8) &                      0.00581(14) &                             0.00516(21) &                       0.0045(4) &                  0.0038(6) &                       0.002940(7) \\
    Si(U23) &                     0.00368(9) &                     0.00371(5) &                      0.00080(13) &                             0.00183(14) &                     0.00101(30) &                  0.0015(6) &                       0.000345(7) \\
    Si(U33) &                     0.00279(8) &                     0.00275(4) &                       0.00421(7) &                             0.00501(10) &                     0.00479(18) &                  0.0056(4) &                       0.004677(5) \\
    Si(U11) &                    0.01179(28) &                    0.01229(17) &                      0.00907(14) &                             0.00815(18) &                     0.00759(34) &                  0.0087(7) &                       0.004906(5) \\
    O(U11) &                    0.01013(33) &                    0.01013(22) &                        0.0372(5) &                               0.0509(9) &                      0.0465(16) &                 0.0491(31) &                      0.021409(27) \\
    O(U22) &                    0.00997(30) &                    0.00997(18) &                      0.02644(33) &                               0.0288(6) &                       0.0224(8) &                 0.0189(15) &                      0.013553(15) \\
    O(U33) &                      0.0130(4) &                    0.01396(23) &                        0.0215(4) &                               0.0190(5) &                       0.0190(9) &                 0.0158(14) &                      0.015211(17) \\
    O(U12) &                    0.01005(30) &                    0.01005(18) &                        0.0318(4) &                               0.0343(7) &                      0.0267(12) &                 0.0269(20) &                      0.011939(19) \\
    O(U13) &                    0.00431(28) &                    0.00431(17) &                      0.00364(24) &                               0.0068(5) &                       0.0072(9) &                 0.0035(15) &                      0.007195(17) \\
    O(U23) &                    0.00294(27) &                    0.00284(16) &                      0.00533(24) &                             0.00593(35) &                       0.0044(6) &                 0.0059(10) &                      0.005919(12) \\
      Si(X) &                    0.47905(19) &                    0.47916(11) &                      0.47663(11) &                             0.47515(12) &                     0.47523(23) &                  0.4759(4) &                       0.470943(6) \\
      O(X) &                    0.42625(25) &                    0.42612(14) &                      0.42082(22) &                             0.41799(28) &                       0.4164(7) &                 0.4196(11) &                      0.415319(14) \\
      O(Y) &                    0.28074(21) &                    0.28072(13) &                      0.27732(15) &                             0.27628(19) &                     0.27540(35) &                  0.2766(6) &                      0.269054(10) \\
      O(Z) &                    0.21318(15) &                     0.21342(9) &                      0.21087(11) &                             0.21034(13) &                     0.20936(26) &                  0.2096(4) &                       0.217546(8) \\
     \qmax &                         16.6 &                         16.6 &                           16.6 &                               16.6 &                          16.6 &                     16.6 &                              26.0 \\
     grid &                       0.189264 &                       0.189264 &                         0.189264 &                                0.189138 &                        0.189264 &                   0.189264 &                           0.12083 \\
       \rw &                       0.295295 &                       0.298084 &                          0.16153 &                                0.146534 &                        0.153905 &                   0.270612 &                           0.13728 \\
  $\chi ^2 _{red}$ &                     120.717794 &                     355.645572 &                       132.124245 &                               72.463849 &                       21.427036 &                  21.665726 &                      11541.295554 \\
\bottomrule
\end{tabular}
\end{adjustbox}
\end{center}
\end{table}
These fits were done on PDFs computed on the Nyquist-Shannon sampling grid, the case for which points in the PDF are as statistically uncorrelated as possible \cite{toby;aca04}, which will result in the most reliable estimates of uncertainties on refined parameters.
The refined values are all very close between the two measurements.  
The most statistically significant variation is with the lattice parameters and the ADPs.
This may indicate an issue with the calibration of one or both of the setups, though the effects are small, with the synchrotron lattice parameters being larger by $\approx 0.01$~\AA. The ADPs of the fitted lab data (P1 and P2) are all larger than from the synchrotron data. As discussed above, this is an effect of the broader peaks in the lab data, which results in overestimated ADPs during fitting. The difference curve between the fitted model and the data in Figure \ref{fig:pdf_fit_quartz}a (P1, without Soller slit) indicates that the beam divergence has a severely negative effect on the PDF of quartz. The origin of this are discussed in section \ref{sec:effects_of_sollerslits}.

\subsection{Comparison of different protocols for \mka data}
We now turn to a comparison of the different \mka data acquisition protocols to obtain the best \mka data for PDF analysis.

\subsubsection{Effect of counting statistics} \label{sec:statistics}

We would like to understand the effects of changing the count-time, and therefore the counting statistics, on the resulting PDF and its information content. We do this by collecting \mka data with identical experiment configuration but different count times and assess the effect on the PDF, and on structural parameters extracted from the PDF. Data was collected using protocol P1 with dwell-times per point of $\deltat = 0.1$, 0.3, and 0.9~s, corresponding to total elapsed experiment times of 1.5, 4.5, and 12.75~hours, respectively.  

\begin{figure}
    \centering
    \includegraphics[width=1.0\textwidth]{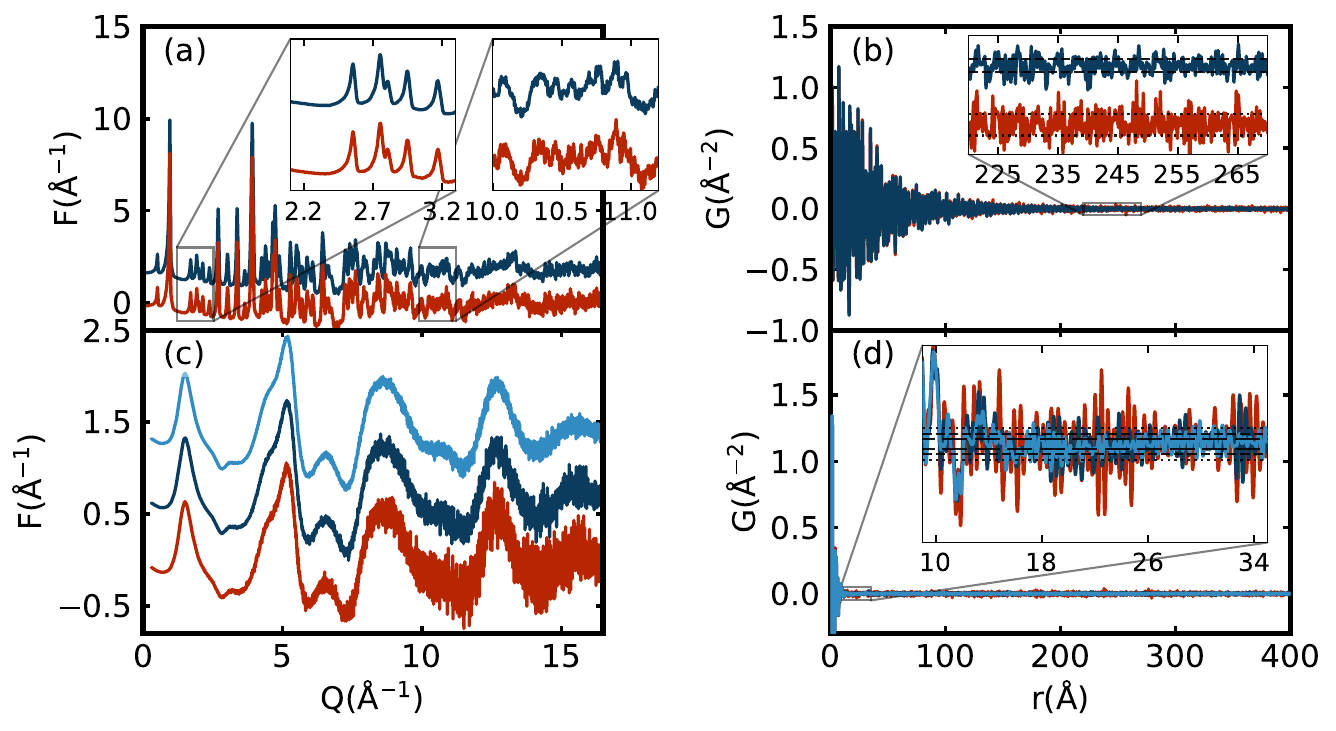}
    \caption{Comparison of the effects of count time on the \mka data where all experiments used protocol P1 but with dwell times per step of $\tstep = 0.1$~s (red), 0.3~s (dark blue) and 0.9~s (pale blue) corresponding to total experiment durations of 1.5, 4.5, and 12.75 ~hours, respectively. (a) and (b) show data from crystalline quartz, and (c) and (d) show data from amorphous silica.  The \fq functions are shown in (a) and (c) and the PDFs, \gr, are shown in (b) and (d). Specific regions of the curves are shown on expanded scales in the insets. Horizontal dashed lines in the insets of (b) and (d) show the root-mean-squared at high-r of $\tstep = 0.1$~s (dots), $0.3$~s (dashes), and $0.9$~s (long dashes).}
     \label{fig:statistics_comparison}
\end{figure}

The longer counting time does result in a reduction of noise in the PDF. This is most easily seen in the \gr of the amorphous silica data in the high-\ir region where the structural signal has gone to zero, and only noise remains in the measured PDF ($9 < \ir < 35$~\AA, inset Figure \ref{fig:statistics_comparison}d). Below $r\sim 15$~\AA, there is a structural signal, but it is gone in the region above this, and the fluctuations are all due to noise in the measurement. 
In this region, we see a reduction in the amplitude of the noise fluctuations with increasing count times, where they are largest in red ($\deltat = 0.1$~s), less so in dark blue ($\deltat = 0.3$~s) and smallest in pale blue ($\deltat = 0.9$~s), consistent with the noise signal being dominated by the counting statistics and therefore reducing in amplitude when the counting time increases. 

If the noise in this region is coming from counting statistics alone, we expect that the rms amplitude of the ripples will decrease as the square-root of the count time.  This is indeed approximately the case. We computed the root-mean-square fluctuations in the $\deltat = 0.1$, 0.3 and 0.9~s data in the region from $300 < r < 400$~\AA~for the silica glass data.  
We get $\deltarms(0.1) = 0.0079$, $\deltarms(0.3) = 0.0048
$, and $\deltarms(0.9) = 0.0027$.
We would expect ratios of $\sqrt{\deltat(1)/\deltat(2)} = \deltarms(2)/\deltarms(1)$.
The results are presented in \tabl{rms}.\\

\begin{table}
    \caption{Expected and actual root-mean-square (rms) fluctuations in the PDF for different ranges of the measurements of the silica glass sample. The "Expected" column contains values that would be expected if the signal is dominated by random counting statistics, $\sqrt{\deltat(1)/\deltat(2)}$, and the "Actual" column is the measured ratio, $\deltarms(2)/\deltarms(1)$, of the rms fluctuations in the signal over the range indicated.}
    \label{table:rms}
    \begin{center}
    \begin{tabular}{cccc}
    \toprule
    \deltat ratio & Range (\iaa) & Expected & Actual \\
    \midrule
     0.3/0.1      & 300-400  & 1.73          & 1.65\\
     0.9/0.1      & 300-400  & 3.00            & 2.93\\
     0.9/0.3      & 300-400  & 1.73          & 1.78\\
     0.3/0.1      & 2-10  & 1.73          & 1.00\\
     0.9/0.1      & 2-10  & 3.00            & 1.00\\
     0.9/0.3      & 2-10  & 1.73          & 1.00\\
     \bottomrule
\end{tabular}
    \end{center}
\end{table}

The actual rms ratios are close to what is expected if they come from random noise.
The actual ratios tend to be slightly smaller than those expected from random counting statistics.
This may be due to some small systematic error in the measured signal.
The noise in the \mka data is counting-statistics limited, which is often not the case in the synchrotron data. In the low-\ir region where the structural signal is strong, the rms ratios are 1.00, which is what would be expected in the opposite limit, where the signal/noise ratio is much greater than one. In this case, the fluctuations in the data are coming from a repeatable structural signal, which doesn't change with count time.

Theoretically, noise in \gr is expected to be roughly constant in~\ir \cite{toby;aca04}, which is one of the attractive aspects of this function \cite{egami;b;utbp12}.
It means that features, for example, in the difference curves of a fit, can be compared at different values of~\ir, and their significance is the same. This is not true for other related pair correlation functions \cite{keenComparisonVariousCommonly2001b}
such as the radial distribution function. Hence, the noise level (peak-to-peak amplitude) in \gr of quartz and silica above $\ir > 200$ and $\ir > 20$~\AA, respectively, are constant in \ir (see dashed lines in insets of Figure \ref{fig:statistics_comparison}b and d).

The fact that the noise level in lab-based PDFs is limited by counting statistics allows us to quantify the noise level in data (by calculating the rms at high \ir).
However, the absolute noise level is not relevant, only the signal-to-noise ratio. We can evaluate the impact of counting statistics on the precision of refined structural parameters in regions with signals. The error calculation at high-r, where there are no structural signals, can be used to estimate uncertainties in refined parameters. In principle, we could propagate the full variance-covariance matrix through the Fourier transform \cite{toby;aca04}, but this is rarely done in practice. Instead, only uncorrelated uncertainties (diagonal matrix elements) are propagated. By computing the rms at high \ir on the Nyquist-Shannon grid, we minimize error correlation as much as possible. A comparison of refined parameters in Table \ref{table:refinement_results} shows that, for well-ordered quartz samples, a 4.5-hour counting protocol (P2(0.3)) provides results comparable to a 25.5-hour protocol (P2(1.8)). However, there's less agreement without using Soller slits (P1 protocol). The uncertainties in P2 are much lower than in P1 and P3 (see numbers in brackets in Table \ref{table:refinement_results}). This indicates that, for precise parameter extraction, better counting statistics and appropriate protocols are essential. 


A challenge when making measurements for PDF analysis is that we measure in~\q but the signals we are interested in are in \ir. So, we would like to build up an intuition of what signals look like in \fq when they give sufficiently good statistics in \gr for our particular experimental setup.
The best reciprocal space function for doing this is \fq because the data have been fully normalized in the way they will be before the Fourier transform.
We show these functions in \fig{statistics_comparison}(a) and (b). 
The way these PDFs (b) appear depends on the step size ($\Delta r$) used to calculate the PDF and can be misleading in a visual comparison.
For example, because we chose a step-size of 0.01~\AA~that oversampled the signal in \gr (see \sect{lab_experiment}) we could resample the data onto a grid that was $2\times$ courser without loss of information.  
This would result in a plot where the apparent noise was reduced by $\sqrt{2}$, but actually, we had not changed the information in the data.
It is, therefore, essential to have a quantitative assessment of the noise level in the PDF. This should be done with PDFs calculated on the Nyquist-Shannon grid. 
Since we are doing PDF analysis, the noise level in the PDF should ultimately govern our decisions.

\subsubsection{Effect of Soller slits}\label{sec:effects_of_sollerslits}

We turn our attention to the effect of Soller slits.
Soller slits reduce the divergence of the incident and scattered beams, improving peak shapes at the cost of lower counting statistics \cite{chearyExperimentalInvestigationEffects1998, chearyAxialDivergenceConventional1998}. Soller slits are often preferred for conventional powder diffraction measurements, especially when low-\q is the scientific focus.
In this region, the signal/noise ratio is high, and there are large gains in line-shape improvements at essentially no cost. For PDF analysis, the high-\q region and high counts are more important, and the use of Soller slits may not always be beneficial.

\begin{figure}
    \centering
    \includegraphics[width=0.65\textwidth]{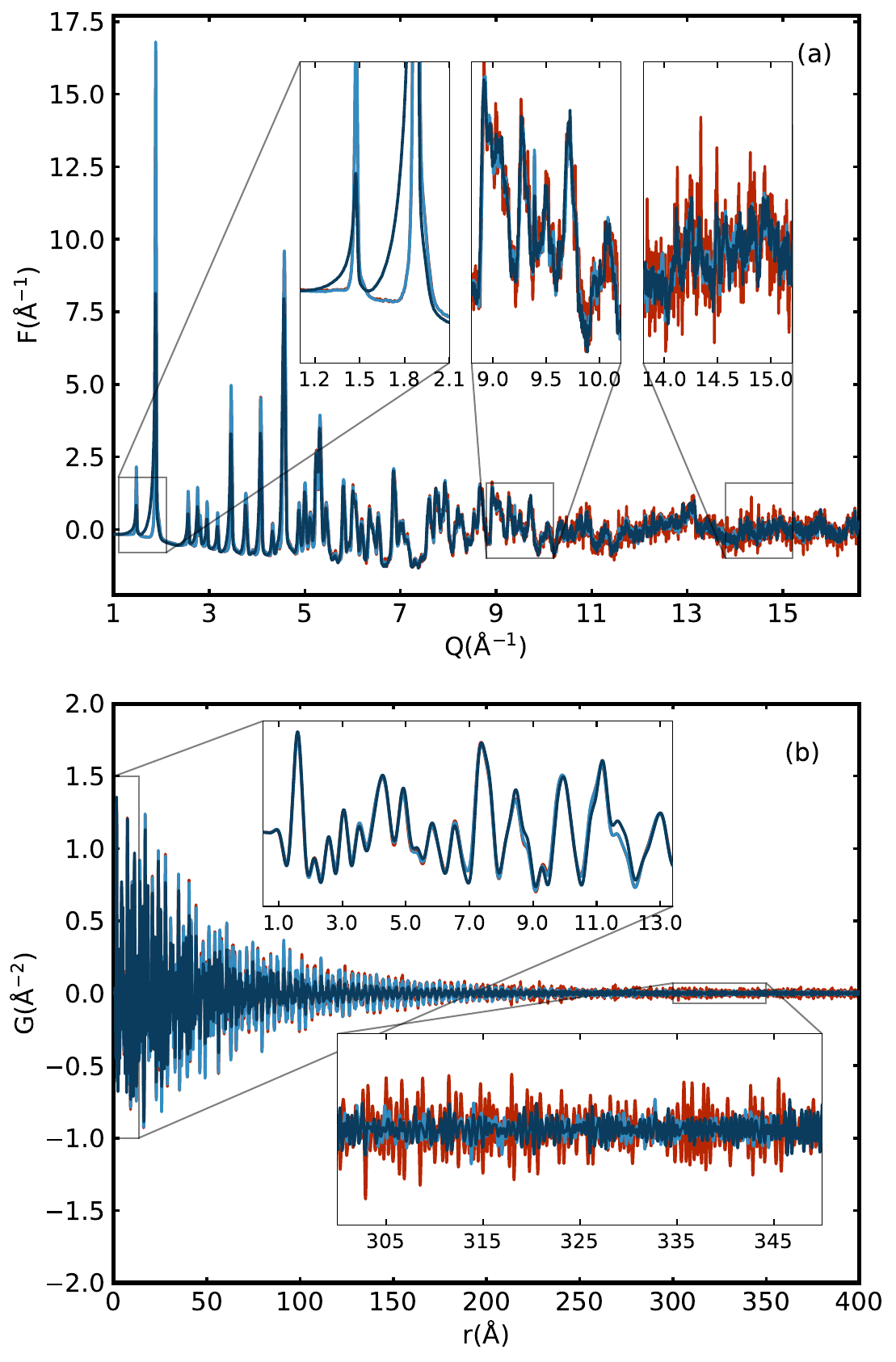}
     \caption{ Comparison of (a) \fq and (b) \gr  of quartz with (red and light blue) and without (dark blue) a 2.5\degree~soller slit.  The data were collected with protocol P1 with $\deltat = 0.3$~s (dark blue), and with protocol P2 with $\deltat = 0.3$~s (red) and $\deltat = 1.8$~s (light blue). Expanded views of regions of the curves are shown in the insets. }
\label{fig:soller_fq_comparison}
\end{figure}

Comparing the \fq of quartz measured with (red and pale blue) and without (dark blue) Soller slits in the diffraction beam (Figure \ref{fig:soller_fq_comparison}a), we see that the Soller slit sharpens the peaks at low-\q (see first inset), but has less of an effect at higher-\q (see middle inset). Comparing the high-\q range, on the other hand, reflects better on the data without Soller slit. In the inset between 14 and 15~\iaa, we see that the noise in the red curve (with Soller slit) is much larger than in the dark blue curve (no Soller slit, same counting time). In fact, we see that with the Soller slits, we have to collect data for around six times longer (25.5~h, pale blue curve) to get approximately the same counting statistics as without the Soller slits (4.5~h, dark blue curve).

The question then arises whether the effect on the Bragg peaks at low-\q, or the reduced counting statistics, has a larger effect on the PDF. The sharper low-\q Bragg peaks results in a PDF where the structural signals extends to much higher-\ir than the data without Soller slit (dark blue, Figure \ref{fig:soller_fq_comparison}b).
Even though for much of the \q-range, the Soller slits do not actually sharpen the Bragg peaks significantly, the sharpening in the low-\q region still results in PDF extending to significantly higher-\ir. On the other hand, in the low-\ir region (\fig{soller_fq_comparison}b upper inset) there is essentially no effect on the amplitude or the width of the PDF peaks due to the addition of the Soller slits (compare dark blue and light blue curves). 

Looking at the fit up to 40~\AA~in \fig{pdf_fit_quartz}(a), we note that, for the \mka data with no Soller slit, there is an evident oscillating signal in the PDF, which was not fitted since it's not coming from the structure (see difference curve).
This likely comes from shifting the intensity away from the correct position of the Bragg peaks in the low-\q region of the diffraction pattern. \cite{coelhoGeneratingAtomicPair2021} have also previously shown the negative effect of axial divergence on resulting PDFs from laboratory measurements, although the main improvement appeared in the high-\ir range due to the applied $K_{\alpha2}$ stripping.
Indeed, the oscillating signal in \gr seems to be a sinusoidal variation of the difference curve with a wavelength of around 3.45~\AA~\fig{pdf_fit_quartz}(a), which would come from an issue in \fq at around $Q=2\pi/3.45 \approx 1.8$~\iaa.  
This is the position of the most substantial low-\q Bragg peak that is significantly affected by the absence of the Soller slits, as shown in \fig{soller_fq_comparison}a (inset).
This would be less of a concern if it did not result in significant differences in refined parameter values.   
We can assess this by comparing columns P1(0.3) (no Soller) and P2(1.8) (with Soller) in \tabl{refinement_results}.
The refined parameters are quite close but do differ beyond the estimated uncertainties.
The results with the Soller slit are closer to those of the synchrotron measurement than are the ones taken without the Soller slit.  
For example, both the synchrotron and the P2 results indicate that the Si ADPs are approximately isotropic, whereas without the Soller slits the fit returns very anisotropic Si ADPs.
The symmetric coordination of Si suggests that it should have rather isotropic ADPs, which further supports the idea that the synchrotron and with-Soller fits are getting it right but not the refinements without the Soller slits.


\begin{figure}
    \centering
    \includegraphics[width=0.8\linewidth]{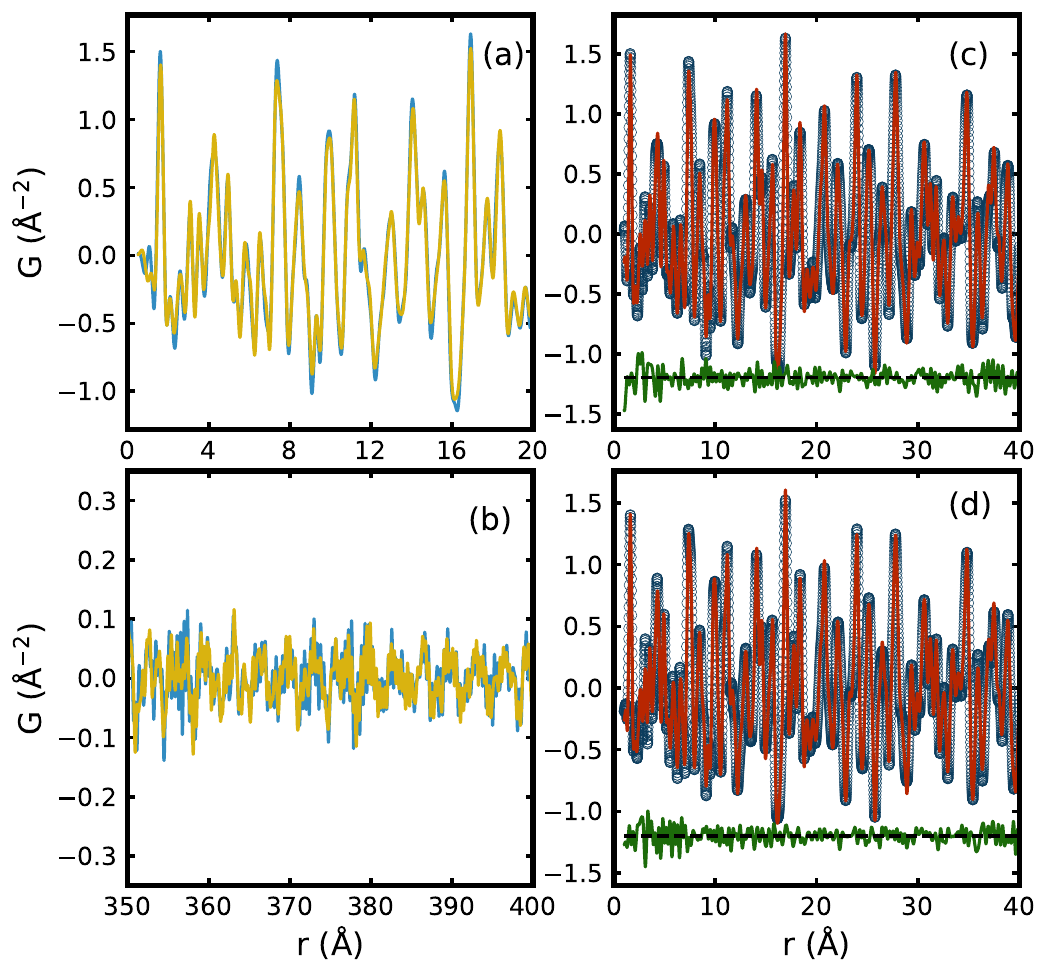}
    \caption{a) Comparison of quartz measured on an Anton Paar XRDynamic 500 measured with protocol P2(1.8) (light blue), and the Anton Paar protocol (yellow) at low-r b) Comparison at high-r, where the noise in the \gr can be used to assess the counting statistics between two measurements. c) quartz fit of the P2(1.8) protocol, and d) fit of the Anton Paar protocol. Specifics of the Anton Paar protocol are listed in Table \ref{tab:ap-protocol}, and the fitting results are listed in Table \ref{tab:anton_paar_params}.}
    \label{fig:antonpaar_fits}
\end{figure}

For diffuse signals, this effect of axial divergence will produce a smaller aberration on the signal, and the importance of better statistics may argue in favor of making measurements with a looser axial divergence (no Soller slits), but for crystalline materials, measurements with Soller slits are clearly preferred. A middle ground between a 2.5\degree Soller slit and no-Soller slit might be favorable in many cases, to reduce the effect of asymmetric peak shapes and increase the counting statistics. For example, Bruker offers a 4\degree axial Soller slits which can be a suitable option for PDF measurements. It is also possible to keep Soller slits installed for the low-angle scans and remove them for the high-angle scans. This may be preferred in some cases since axial divergence's negative effect mainly affects the low-angle peaks, where we want to improve the lineshapes as much as possible. We have tested such a case and compared PDF obtained on the XRDynamic 500 \cite{antonpaar} using the P2(1.8) protocol and a protocol developed by Anton Paar, where the Soller slits are removed for the high-angle scans (see Table \ref{tab:ap-protocol} for specifics). The PDFs from the two protocols are essentially the same in the low-\ir region (Figure \ref{fig:antonpaar_fits}a) and the noise level is similar too, despite the much lower counting time in the Anton Paar protocol (Figure \ref{fig:antonpaar_fits}b). The fitting of the obtained PDFs shows that the Anton Paar protocol yields a PDF of the same quality in 8 hours compared to 25.5 hours for the P2(1.8) protocol (Figure \ref{fig:antonpaar_fits} c and d). No signs of an oscillating signal can be observed in the difference curve, which means that removing the Soller slit for the high-\q scans has no negative impact on the resulting PDF. The refined parameters, including uncertainty estimation, are listed in Table \ref{tab:anton_paar_params}, and details to the Anton Paar protocol are listed in Table \ref{tab:ap-protocol}. 

\subsubsection{Effect of beam length on the sample}

\begin{figure}
    \centering
    \includegraphics[width=0.5\textwidth]{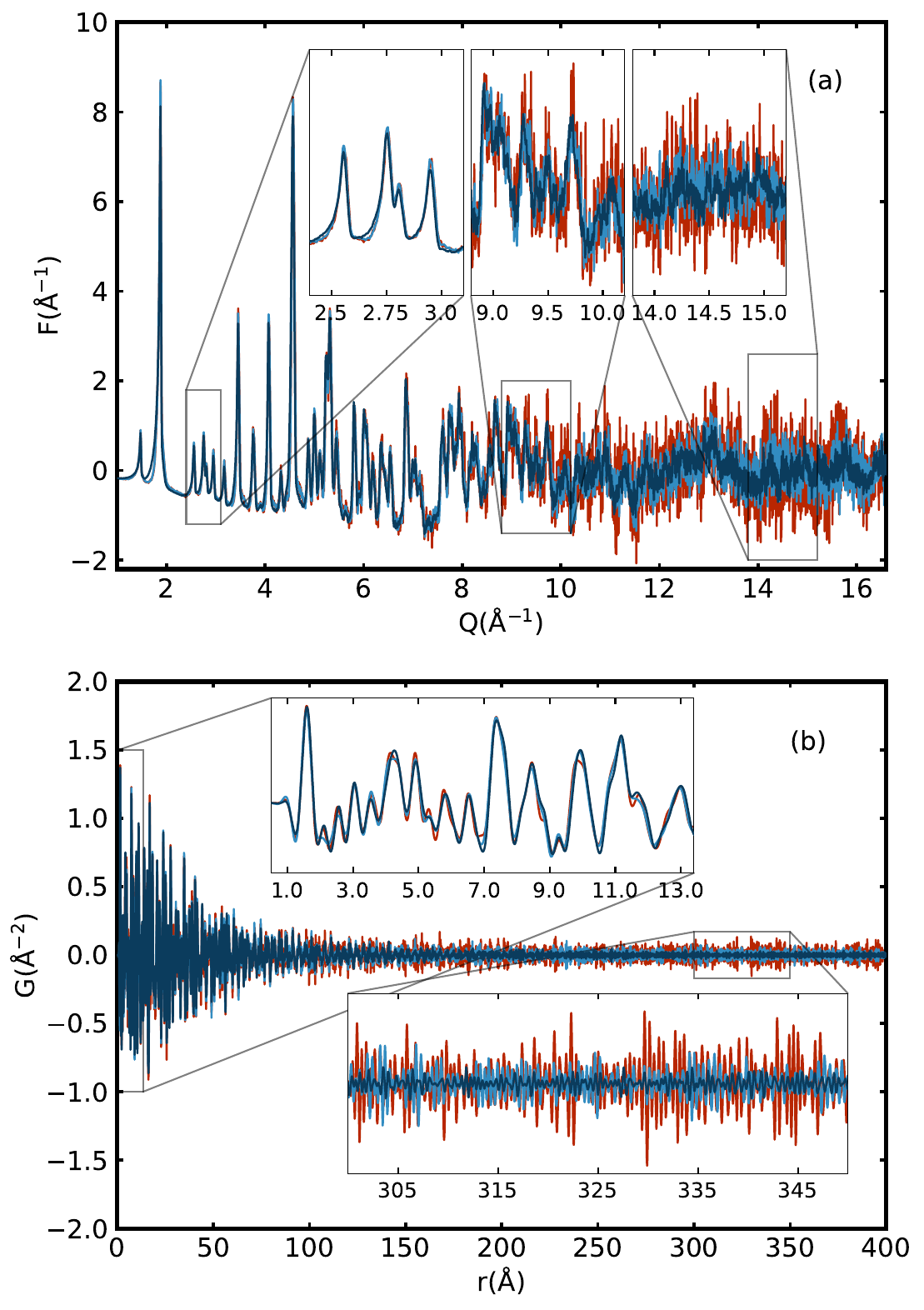}
     \caption{ Comparison of (a) \fq and (b) \gr  of quartz with a 15~mm illumination length on the sample (red and light blue) and with 25~mm illumination length (dark blue).  The data were collected with protocol P3 with $\deltat = 0.1$~s (red) and $\deltat = 0.3$~s (light blue) and P1 with $\deltat = 0.1$~s (dark blue), and with protocol. Expanded views of regions of the curves are shown in the insets.}
     \label{fig:length_illuminated_comparison}
\end{figure}

We can also look at the effect of using a mask to limit the length of the incident beam along the capillary. 
In the current case, a mask was crafted using lead tape to restrict the beam length on the sample from 25~mm to 15~mm. This masks out some of the out-of-plane beam divergence.
The results are shown in  \fig{length_illuminated_comparison}.

The results of this test are very similar to the case of the axial Soller slits, which limit the out-of-plane divergence of the beam.
Inserting the Pb-mask results in slightly sharper Bragg peaks in the low-\q region of the diffraction pattern but not in the mid-high~\q regions and results in noisier data for the same count time.
In the low-\ir region of the PDF, the peak shapes and amplitudes are essentially unaffected by the removal of the Pb-mask but are slightly more noisy.
The sharpening of the Bragg peaks is less acute than in the case of the axial Soller slits, and so the extension of the signal in the high-\ir region is less pronounced.

The effect on refined parameters can be assessed by comparing columns P3(0.3) with P1(0.1).  
The two result in quite comparable refinements, though the P3(0.3) protocol gave results slightly closer to the synchrotron and Soller slit cases.

Overall, the trade-off between count time and resolution seems to favor the use of the full out-of-plane divergence (no axial Soller slits or Pb-mask) for PDF measurements. 
However, limiting beam divergence at the cost of longer count-times seems favorable for the most accurate measurements on highly crystalline materials. 

\section{Multiplicative correction}\label{sec:mult_corr}
\subsection{Absorption effects}\label{sec:abs}

Here, we explore the modification to the intensity due to sample absorption for our capillary setup.
Many solutions to absorption corrections in capillary geometries exist, mainly for reliable refinement of PXRD data \cite{walkerEmpiricalMethodCorrecting1983, sabineAnalyticalExpressionsTransmission1998, coelhoCapillarySpecimenAberration2017, sulyanovSpatialDistributionAbsorption2012}. Our solution is designed to be used with \getx to correct raw data before data reduction.
In general, the measured intensity is proportional to the effective sample volume probed by the x-ray beam, assuming a uniform sample density.
In the absence of significant x-ray absorption by the sample this effective sample volume is just the volume of the physical region of the sample that is illuminated by the x-rays.  
For common geometries used in \rapdf measurements at synchrotrons, cylindrical samples, or flat pill-shaped samples perpendicular to the beam with the detector behind the sample, the illuminated volume is not sattering-angle dependent and therefore not \q~dependent.
In such cases, no multiplicative corrections are needed, and \getx can reliably produce PDFs from that data.

When absorption by the sample is significant, the effective volume deviates from the physical volume probed because there are different path-lengths that the x-rays have to take to reach different sub-regions, or voxels, of that physical volume.
The measured scattered intensities from the different voxels then become weighted by how many x-rays are absorbed by the sample on the path the x-rays take to reach that voxel, and then the path to exit the sample again to reach the detector.
This results in an angle dependence on the effective volume of the sample, which requires an angle-dependent multiplicative correction to the measured intensities.
In the forward-scattering directions, this angle dependence is weak, which further helps the approximation of no multiplicative correction for the \rapdf geometry.
However, when intensities are collected over wide angular ranges, the effects can be large. Part of the upturn in the \mka data evident in \fig{demo_abs_corr} will originate from this effect because we expect the effective sample volume to be larger in backscattering because more of the sample voxels can be accessed by short path-lengths through the sample.


\subsection{Effective volume correction}

To correct our measured signal $I_m$ for sample absorption, we have to compute the x-ray attenuation $A_v (\tth)$ for each voxel $v$ in the cappillary and as a function of \tth:
\begin{equation}
A_v(\tth) = e^{-\mu_s\ell_v(\tth)}
\end{equation}
where $\mu_s$ is the sample linear x-ray attenuation coefficient, where we assume a macroscopically uniform sample so it does not depend on the location of the voxel.
Using this, we can calculate the effective volume $v_e$, which includes the angle-dependent x-ray attenuation
\begin{equation}
    v_e = \frac{1}{N_v}\sum_v e^{-\mu_s\ell_v(\tth)}\label{eq:v-e-const-pixel},
\end{equation}
Where the sum is over all voxels in the sample, and $N_v$ is the number of voxels. Our corrected intensity $I_c$ then becomes:
\begin{align}
    I_c &= \frac{I_m}{v_e} \\
        &= \frac{N_vI_m(\tth)}{\sum_v e^{-\mu_s\ell_v(\tth)}}\label{eq:ic-final}
\end{align}
More details about the calculation of the effective volume and the derivation for eq. \ref{eq:ic-final} can be found in \textbf{section 2} of the \textit{supplementary information}.
\begin{figure}
    \centering
    \includegraphics[width=0.6 \linewidth]{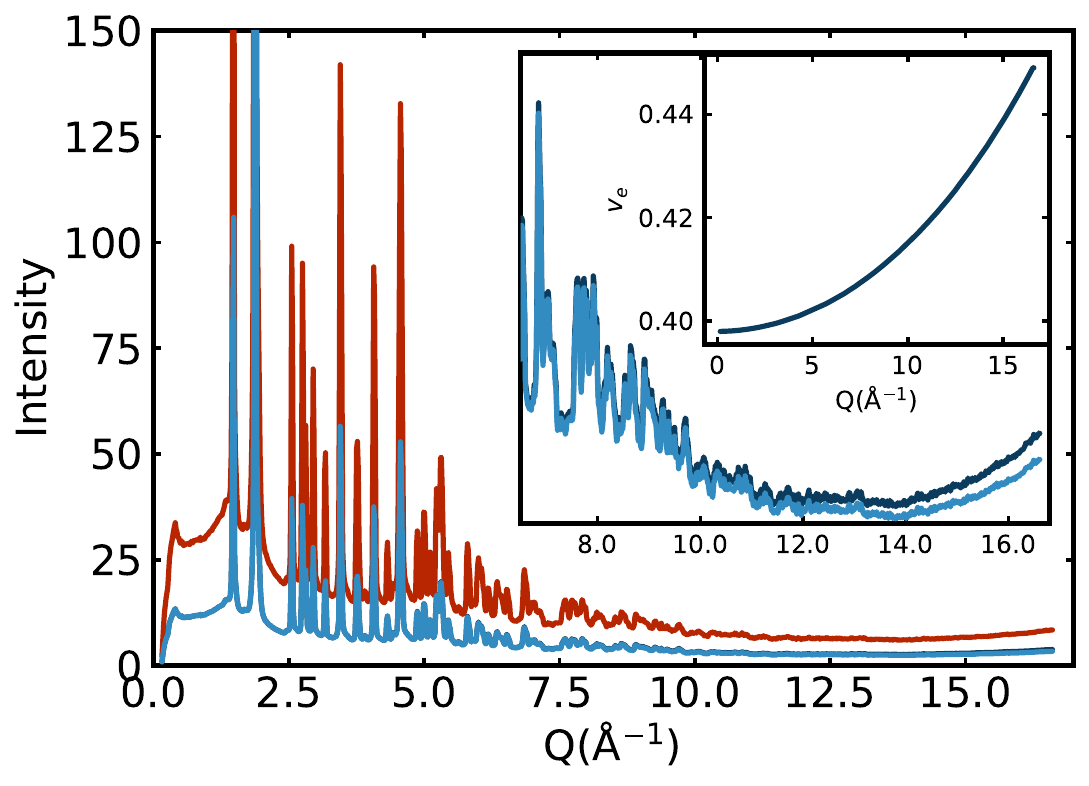}
    \caption{Quartz \mka diffraction pattern with and without correction for absorption.  The raw, uncorrected, signal is dark blue and the absorption corrected signals is red.  A version of the absorption corrected signal that has been scaled down to lie on top of the raw signal is shown in light blue.  The high-\q behavior of the raw and scaled-corrected curves are shown on an expanded scale in the inset.  In the second inset we show the effective volume, $v_e$, from the absorption correction for a 1.5~mm diameter tube of quartz with and \mka radiation.}
    \label{fig:abscorr}
\end{figure}

In \fig{abscorr} we show the angle-dependence of the effective volume,  $v_e(\tth)$, for the case of our \mka experiments on the silica samples.
This was computed for the sample composition \ch{SiO2} and assuming a packing fraction of 0.66 resulting in $\mu_s = 7.3$~cm$^{-1}$, and a 1.5~mm ID capillary (\mud = 1.1, where $D$ is the diameter of the tube).
The absolute value of $v_e$ for this case is a little less than 0.5, with a \q-dependence from low-\q to high-\q of a little less than 0.05.  
Thus, for \mud of the order of unity, the effects on the resulting PDF are small.

\subsection{Effect of the $v_e$ correction on the measured signal}

\fig{abscorr} shows the raw and corrected diffraction curves for our \mka quartz.
Dividing by $v_e$ increases the overall scale factor of the corrected signal.  
This is seen in the main panel where the raw signal is shown in dark blue and the corrected signal in red.  
Since the \getx algorithm is scale agnostic, this overall change in scale will not result in a change in the resulting PDF. 
However, as we have discussed, there is also \iq-dependence to the absorption correction where the effective volume is larger in the back-scattering region.
When we divide by $v_e$ we will therefore preferentially scale down the high-\iq signal compared to the low-\q signal.  This is shown in the inset where we have taken the corrected curve in red and scaled it down so that it lies on top of the raw signal in the low-r region. 
We see that in the high-\q region the absorption corrected signal increases by less.
We see from the plot of $v_e$ that the \q-dependence of the curve is an order of magnitude less than the absolute value.  
It increases from 0.4 to $\sim 0.45$.

The comparison of refined parameters with and without the absorption correction can be assessed by comparing the columns P2(1.8) and P2(1.8)~``Abs" in \tabl{refinement_results}. The fit of the corrected data has a slighly lower \rw (0.147 compared to 0.162), and the refined lattice parameters are closer to the values from the synchrotron data. However, the effects are overall small due to the moderate \mud of 1.1. It is expected that samples with larger \mud would have a stronger effect, and applying an absorption correction would result in more reliable PDFs in such cases.

\section{Protocol recommendations}\label{sec:recommendations}

As discussed, there is often a trade-off between optimal counting statistics and reasonable acquisition time. Shorter times are generally preferred, especially for users frequently performing PDF analysis. Different protocols should be considered for crystalline and amorphous samples. For crystalline samples, low-angle peak lineshape significantly affects the PDF, while amorphous samples, having lower scattering intensities, are best measured without Soller slits.\\

\begin{itemize}
    \item \textbf{Estimate \mud}. A capillary diameter between 0.5 and 1.5~mm (ID) is preferred for convenience and resolution.  Pick the diameter to get a \mud as close to one as possible. If $\mud > 1$, an absorption correction may need to be applied.  If $\mud > 6$ or so, you are unlikely to get workable data.
    \item \textbf{Define your setup.} For crystalline samples, install an axial Soller slit. There are several options to consider: 
    \begin{itemize}
        \item i) Install an axial Soller slit for the whole \tth range. 2.5\degree openings are standard equipment but reduce the counting statistics significantly. If available, larger openings like a 4\degree Soller slit is advised.
        \item ii) Install an axial Soller slit for low angle scans and remove them for high angles scans, as shown in Table \ref{tab:ap-protocol}.
    \end{itemize}
    \item \textbf{Set up scans.} As shown in Figure \ref{fig:tstep}, beak up the available angular range ($\tth_{min}-\tth_{max}$) into 4-5 scans. One can either employ a staircase-counting-time (SCT) protocol as outlined in Table\ref{table:staircase}, or some other variable counting time (VCT) protocol outlined in Table \ref{tab:ap-protocol}. The individual scans have to be combined into one PXRD pattern. This can be done typically by the instrument software. E.g. for the SCT, define a base counting time per step \deltat (see Table \ref{table:staircase}), and double it for each consecutive higher-Q scan. A \deltat of 0.3 seconds for the first scan should suffice for crystalline samples. For amorphous samples, higher counting times may be necessary (\deltat of 0.9 or 1.8). This depends a lot on the specifics of the instrument and detector, and decisions on \deltat should be made based on the noise level in \fq and \gr, as described in this paper.
    \item \textbf{Apply absorption correction (optional).} Depending on the \mud, an absorption correction of the raw XRD data may improve the resulting PDF. We provide a freely available Python package \textit{labpdfproc} to process laboratory XRD data for PDF analysis. The package can be found here: \url{https://github.com/diffpy/diffpy.labpdfproc}. All that is required for user input is an estimation of \mud. The package returns a corrected PXRD pattern, which can be used for subsequent data reduction. Several tools are available to calculate the \mud of a given sample, for instance, here: \url{https://11bm.xray.aps.anl.gov/absorb/absorb.php}.
    \item \textbf{Evaluate the data.} Use \getx or any other tool to obtain \fq and \gr. As discussed in this paper, the high-r range can be used to evaluate the noise level between different measurements. It is advised to calculate the PDF both on a 0.01~\AA~grid and on the Nyquist-Shannon grid ($\Delta r = \frac{\pi}{\qmax}$). We also provide our DiffPy-CMI python script as part of the \textit{labpdfproc} package, which can be used to fit and compare samples measured with different acquisition protocols or to compare corrected and non-corrected data. The script uses the rms calculated at high-\ir for error propagation, so the input PDFs have to be calculated up to 500~\AA. PDFs on a 0.01 grid are sufficient since the script recalculates the PDFs on an NS grid automatically.
\end{itemize}
\section{Conclusion}

We have carried out a systematic assessment of different protocols for measuring data on a laboratory \mka source.
We present a particular variable counting scheme, the staircase count time scheme, that results in preferential counting in the high-\q region where the signal/noise ratio is worst for PDF data.
We show that for crystalline materials and PDF modeling, the \mka data gave comparable results to the synchrotron measurement. The uncertainties in the \mka data were shown to be dominated by counting statistics, whereas the synchrotron data fits are model-limited, and uncertainty estimates do not reflect the true accuracy.
For PDF model refinements on crystalline materials, limiting the axial divergence, in our case with the use of Soller slits in the secondary flight-path, gave more accurate results. We present strategies, such as the removal of axial Soller slits for higher-angle scans, in order to improve counting statistics whilst preventing the negative effects of axial divergence at low angles.
We report additional corrections that may be introduced in the data reduction process to account for sample absorption.
For our case, with a $\mud \sim 1.1$ computed with a packing fraction of 0.66, the effects were small. We emphasize that for samples with a larger \mud, an absorption correction may be required to obtain reliable PDFs from laboratory diffractometers.
It is recommended to assess sample capillary thickness to obtain a \mud close to unity or below at the energy of the laboratory x-ray source.
However, for more absorbing samples, we present software that may be inserted into the data reduction steps to correct for these effects.


\section{Acknowledgements}

Work in the Billinge group was supported by a grant from 3M Corporate
Research Analytical Laboratory. Synchrotron x-ray PDF measurements were conducted on beamline 28-ID-1 of the National Synchrotron Light Source II, a U.S. Department of Energy (DOE) Office of Science User Facility operated for the DOE Office of Science by Brookhaven National Laboratory under Contract No. DE-SC0012704. Till Schertenleib acknowledges support from the Swiss National Science Foundation under grant number 200021\_188536. The authors gratefully thank Anton Paar GmbH for carrying out measurements of quartz on their XRDynamic 500 diffractometer.

\section{Data availability}

Data used for all the plots in the manuscript are available on Zenodo at \url{https://doi.org/10.5281/zenodo.11060384}

\bibliographystyle{unsrt}
\bibliography{billinge-group-bib,bg-pdf-standards,ts_mo_data_reduction, web_citations.bib}

\begin{thebibliography}{10}
\bibitem{whittakerXVIIIOnFunctionsWhich1915}
E.~T. Whittaker.
\newblock {{XVIII}}.---{{On}} the {{Functions}} which are represented by the {{Expansions}} of the {{Interpolation-Theory}}.
\newblock {\em Proceedings of the Royal Society of Edinburgh}, 35:181--194, January 1915.
\end{thebibliography}

\newpage

\section{Supplementary Information}

\section{Anton Paar XRDynamic 500 measurements}

Quartz was measured on an Anton Paar XRDynamic 500 using the staircase-counting time (SCT) P2(1.8) up to 140\degree~\tth and a 2.86\degree~axial Soller slit installed in the diffraction beam. For comparison, a modified variable counting time protocol by Anton Paar was used as shown in Table \ref{tab:ap-protocol}. The scans go up to 160\degree~\tth and have the Soller slit installed for the low angle peaks up to 80\degree~\tth, and removed for the scans from 80-160\degree~\tth. The resulting PDF was fitted and the refined parameters including error propagation and uncertainty estimations are listed in Table \ref{tab:anton_paar_params}.\\
\\

\begin{table}[H]
\begin{center}
     \caption{Instrument specifics for the Anton Paar XRDynamic 500 and the counting protocol for the Anton Paar protocol.}
    \begin{tabular}{ll}
    \toprule
         measurement geometry & transmission, capillary/Debye-Scherrer \\
         X-ray tube & \mka, 55kV and 50 mA (line focus) \\
         Optics & Focusing beam mirror \\
         Detector & Pixos 2000 CdTe \\
         Step size & 0.02\degree \\
         Divergence slit & 0.1\degree \\
         Beam mask & 10 mm \\
         Scan range (\tth) & 2-41\degree, 40-81\degree, 80-121\degree, 120-160\degree \\
         counting time per step ($s^{-1}$) & 1.1 (scan 1), 2.08 (scan 2), 3.66 (scan 3), 7.3 (scan 4) \\
         Axial soller slits &  2.86\degree (scan 1), 2.86\degree (scan 2), none (scan 3), none (scan 4) \\
    \bottomrule
    \end{tabular}
    \label{tab:ap-protocol}
\end{center}
\end{table}

\begin{table}[H]
\begin{center}
\caption{Comparison of quartz measured on an Anton Paar XRDynamic 500 with \mka source, using protocol P2(1.8) with Soller slits, and an Anton Paar protocol which has the Soller slits installed for low angle scans and removed for high angle scans. The measurement specifics for the latter are shown in Table \ref{tab:ap-protocol}.}
\label{tab:anton_paar_params}
\begin{tabular}{lll}
\toprule
Parameter & P2(1.8) & Anton Paar protocol \\
\midrule
       scale &                                         0.294(8) &                              0.2715(35) \\
    \qdamp &                                    -0.0(2.6)e+09 &                               -0(5)e+12 \\
   \qbroad &                                         0.04(14) &                              0.0350(29) \\
   $\delta _2$ &                                          -0.3(8) &                                1.26(30) \\
        a &                                         4.912(4) &                               4.9145(5) \\
        c &                                       5.4030(12) &                              5.4053(11) \\
    U22\_0 &                                        0.0038(5) &                              0.0032(13) \\
    U23\_0 &                                        0.0006(6) &                              -0.0000(7) \\
    U33\_0 &                                        0.0040(4) &                               0.0025(6) \\
    U11\_0 &                                        0.0072(6) &                              0.0060(11) \\
    U11\_3 &                                       0.0101(12) &                                0.010(4) \\
    U22\_3 &                                        0.0100(9) &                              0.0100(12) \\
    U33\_3 &                                       0.0153(14) &                              0.0204(23) \\
    U12\_3 &                                        0.0101(9) &                              0.0091(26) \\
    U13\_3 &                                        0.0063(8) &                              0.0059(13) \\
    U23\_3 &                                        0.0068(9) &                              0.0065(11) \\
      x\_0 &                                        0.4729(6) &                               0.4713(7) \\
      x\_3 &                                        0.4202(9) &                              0.4206(15) \\
      y\_3 &                                        0.2699(8) &                              0.2693(12) \\
      z\_3 &                                        0.2201(7) &                              0.2186(11) \\
     \qmax &                                             16.6 &                                    17.4 \\
     grid &                                         0.189253 &                                0.180551 \\
       \rw &                                          0.17879 &                                0.156573 \\
  $\chi_{red}^{2}
  $ &                                         4.663061 &                                3.348481 \\
\bottomrule
\end{tabular}
\end{center}
\end{table}

\section{Derivation of effective volume correction}

To determine the effective sample volume, \esv, for a capillary geometry we first determine the x-ray path length into and out of each voxel at each scattering angle, \tth, $\ell_v(\tth)$.
We can then determine the absorption for the $v^\mathrm{th}$ voxel, $A_v(\tth)$, from
\begin{equation}
A_v(\tth) = e^{-\mu_s\ell_v(\tth)}
\end{equation}
where $\mu_s$ is the sample linear x-ray attenuation coefficient, where we assume a macroscopically uniform sample so it does not depend on the location of the voxel.
The normalized scattering is then obtained by dividing the measured intensity by the effective sample volume given by
\begin{equation}
    V_e = \sum_v A_v \Delta V_v,
\end{equation}
where $\Delta V_v$ is the volume of the $v^\mathrm{th}$ voxel and the sum is over all the voxels in the illuminated region of the sample.
The volume-corrected intensity, $i_c$ is then given by 
\begin{equation}
    i_c(\tth) = \frac{I_c(\tth)}{V} = \frac{I_m(\tth)}{V_e(\tth)},
\end{equation}
for a measured intensity $I_m$. To get $v_e(\tth)$, the angle-dependence of $V_e(\tth)$, we divide $V_e(\tth)$ by the physical illuminated volume, $V = \sum_v \Delta V_v$,
\begin{align}
    v_e &= V_e/V \\
        &= \frac{\sum_v e^{-\mu_s\ell_v(\tth)}\Delta V_v}{\sum_v \Delta V_v}\label{eq:v-e}.
\end{align}
If all the voxels are the same size, $\sum_v \Delta V_v = N_v \Delta V$ and we get
\begin{equation}
    v_e = \frac{1}{N_v}\sum_v e^{-\mu_s\ell_v(\tth)}\label{eq:v-e-const-pixel},
\end{equation}
and our corrected intensity is
\begin{align}
    I_c &= \frac{I_m}{v_e} \\
        &= \frac{N_vI_m(\tth)}{\sum_v e^{-\mu_s\ell_v(\tth)}}\label{eq:ic-final}
\end{align}

\end{document}